\documentclass[12pt]{article}
\usepackage{amsmath}
\usepackage{amssymb}
\usepackage{stmaryrd}
\usepackage{color}
\usepackage{ wasysym }
\usepackage{color}

\newcommand{\supp}{\mathop{\mathrm{\supp}}}

   \newtheorem{theorem}{Theorem}[section]
   \newtheorem{lemma}[theorem]{Lemma}
   \newtheorem{proposition}[theorem]{Proposition}
   \newtheorem{corollary}[theorem]{Corollary}

\newcommand{\ben}{\begin{enumerate}}
\newcommand{\een}{\end{enumerate}}

\newcommand{\bt}{\begin{theorem}}
\newcommand{\et}{\end{theorem}}
\newcommand{\bl}{\begin{lemma}}
\newcommand{\el}{\end{lemma}}
\newcommand{\bc}{\begin{corollary}}
\newcommand{\ec}{\end{corollary}}
\newcommand{\bp}{\begin{proposition}}
\newcommand{\ep}{\end{proposition}}
\newcommand{\br}{\begin{remark}}
\newcommand{\er}{\end{remark}}

\newcommand{\bpf}{\begin{proof}}
\newcommand{\epf}{\end{proof}}

\newcommand{\be}{\begin{equation}} 
\newcommand{\ee}{\end{equation}}
\newcommand{\beq}{\begin{eqnarray}}
\newcommand{\eeq}{\end{eqnarray}}
\newcommand{\ba}{\begin{array}}
\newcommand{\ea}{\end{array}}
\newcommand{\bi}{\begin{itemize}}
\newcommand{\ei}{\end{itemize}}

\newcommand{\comm}[1]{}

\newcommand \qed {\hskip 6pt\vrule height6pt width5pt depth1pt \bigskip}

\newfont{\msbm}{msbm10 scaled\magstep1}
\newfont{\msbms}{msbm7 scaled\magstep1} 

   \newenvironment{proof}[1][Proof]{\begin{trivlist}
   \item[\hskip \labelsep {\bfseries #1}]}{\end{trivlist}}

   \newenvironment{remark}[1][Remark]{\begin{trivlist}
   \item[\hskip \labelsep {\bfseries #1}]}{\end{trivlist}}

   \comm{\newcommand{\qed}{\nobreak \ifvmode \relax \else
      \ifdim\lastskip<1.5em \hskip-\lastskip
      \hskip1.5em plus0em minus0.5em \fi \nobreak
      \vrule height0.75em width0.5em depth0.25em\fi}}

 \numberwithin{equation}{section}

\usepackage{fancyhdr,amsfonts, amsmath}

\begin{document}

\title{Griffiths inequalities for non-interacting rotors}
\author{Ira Herbst}
\date\today

\maketitle


ABSTRACT: We prove Griffiths inequalities for spins in $n \ge 2$ dimensions with no interaction.

\section{Introduction}

An old open problem is to prove correlation inequalities $E fg \ge (E f) (E g) $ where the expectation is in the ferromagnetic measure $$E f = \int  f(\sigma_1,...,\sigma_N) e^{\sum_{i,j} f_{i,j}\sigma_i\cdot \sigma_j} d\sigma_1 \cdots d\sigma_N/Z$$.

Here $\sigma_i \in \mathbb{S}^{n-1}, n\ge 2$, $d\sigma_i$ is normalized Lebesgue measure on $\mathbb{S}^{n-1}$, and $Z$ is chosen so that $E$ is expectation in a probability measure.  The fact that the measure is ferromagnetic means $f_{i,j} \ge 0$.  The open problem is to show (or disprove by example) that $E fg \ge (E f) (E g) $ if both $f$ and $g$ are of the form $\prod_{1\le i< j\le N} (\sigma_i\cdot \sigma_j)^{n_{i,j}}$.  Ferromagnetic spin inequalities were proved for $n=1$ in special cases by Robert Griffiths \cite{Gr1, Gr2} and more generally by D. Kelly and Seymour Sherman {\cite{KS}.  In  $n=2$ they were proved by Jean Ginibre \cite{G}.  Ginibre's proof relied on even stronger inequalities which he proved for $n=2$.  But these were shown to be false in general for $n > 2$  by Garrett Sylvester \cite {GS} with explicit counterexamples.  There are correlation inequalities known in dimensions three and four with ferromagnetic interaction but to my knowledge not any of the correlation inequalities mentioned above.  See the review article \cite{MP} of James Monroe and Paul Pearce.

In the next section we prove the correlation inequalities in the case $f_{i,j} = 0$ all $i,j$.  This is of course physically uninteresting but mathematically non-trivial.  They have been proved by a completely different method by Bernhard Baumgartner \cite {B}.

I proved these results in 1979-80 but never published the proof. Now in my retirement years I thought this method of proof should see the light of day to perhaps spur additional research.  

In the last section using the same semigroup technique we prove Griffiths inequalities for Gaussian spins with a ferromagnetic interaction.  This actually has a trivial proof (see for example \cite {GS}) using a change of variable and Isserlis' theorem \cite{I} (or Wick's theorem \cite{W}) for Gaussian integrals.  But we give a semigroup proof, again to possibly spur additional research.

\section{The inequalities}
Let $\mathcal{E}$ be the set of linear combinations of monomials of the form $\prod_{1\le i< j\le N} (\sigma_i\cdot \sigma_j)^{n_{i,j}}$ with positive coefficients.  Here $\sigma_i \in \mathbb{S}^{n-1}, n \ge 2$, $n_{i,j} \ge 0$. Define
\be \label{Edef}
E f = \int f(\sigma_1,\cdots,\sigma_N) d\sigma_1 \cdots d\sigma_N
\ee

 where $d\sigma_j$ is normalized Lebesgue measure on $\mathbb{S}^{n-1}$.  

\bt
Suppose $f,g \in \mathcal{E}$.  Then the first and second Griffiths inequalities are valid:  $E f \ge 0$ and $E fg \ge Ef Eg$, where $E$ is given by (\ref{Edef}).
\et

\bpf
We use a method devised by Loren Pitt \cite{P} and used by Herbst and Pitt \cite {HP} to prove results related to the FKG inequalities.  The idea is to find a self adjoint operator $A \le 0$ with the function $1$ the non-degenerate normalized ground state of $-A$ with $A 1 = 0$ so that for real $f$ and $g$ in $L^2$, $E f e^{tA} g = (f,e^{tA}g) \rightarrow (f,1)(1,g)$ as $t \rightarrow \infty$.  Thus if one can show $(d/dt) E f e^{tA} g \le 0$ for particular functions $f$ and $g$ we have $E fg \ge (E f) (E g) $ .

We take $A = \Delta = \sum_{i=1}^N \Delta_i$ where $\Delta_i$ is the spherical Laplacian on the $i$th sphere.    The main step is to show that if $f \in \mathcal{E}$ then $e^{t\Delta}f $ is a limit of functions in $\mathcal{E}$.  Since with $f, g \in \mathcal{E}$, $$(d/dt)Ef e^{t\Delta} g = - \int \nabla f \cdot \nabla e^{t\Delta} g  d\sigma_1 \cdots d\sigma_N, $$
it then remains to show that  $\int (\nabla f \cdot \nabla h)  \prod_{i=1}^N d\sigma_i \ge 0$ for $f,h \in \mathcal{E}$.  Here $\nabla f \cdot  \nabla h = \sum_{i=1}^N \nabla_i f\cdot \nabla_i h$ and $\nabla_i$ is the gradient on the $i$th sphere.  Thus for example $\nabla_i \sigma_i \cdot a = a - (a \cdot \sigma_i ) \sigma_i  = (I - P_{\sigma_i}) a $ where $P_{\sigma_i}$ is the orthogonal projection onto multiples of $\sigma_i$. We find 
$$\nabla_1(\sigma_1 \cdot \sigma_i) \cdot \nabla_1 (\sigma_1 \cdot \sigma_j) =  \sigma_i \cdot (I- P_{\sigma_1}) \sigma_j.$$
Thus to show $\int (\nabla f \cdot \nabla h)  \prod_{i=1}^N d\sigma_i \ge 0$ for $f,h \in \mathcal{E}$ it suffices to show for non-negative integers $n_{lk}$  that the integral 

\be \label{integral}
J = \int  [\sigma_i \cdot (I- P_{\sigma_1}) \sigma_j ]\prod_{l,k}  ( \sigma_l\cdot\sigma_k)^{n_{lk} }d\sigma_1 d\sigma_2 \cdots d\sigma_N 
\ee
is non-negative.
To see that $J \ge 0$, we first note that by rotation invariance we can omit the integral over $\sigma_1$ and take $\sigma_1 = (1,\cdots,0)$ without changing the value of (\ref{integral}).  Thus we are left with sums with positive coefficients of integrals of the form ($(\sigma_i)_m$ is the $m$th component of the vector $\sigma_i$)
$$ \int \sum_{m=2}^n (\sigma_i)_m(\sigma_j)_m (\sigma_{l_1})_{m_{l_{1}}}  (\sigma_{l_2})_{m_{l_{2}}} \cdots  (\sigma_{l_M})_{m_{l_{M}}} d\sigma_2 \cdots d\sigma_N$$
where $l_p >1$ or more generally sums with positive coefficients of integrals of the form 

 $$ I = \int(\sigma_{l_1})_{m_{l_{1}}}  (\sigma_{l_2})_{m_{l_{2}}} \cdots  (\sigma_{l_M})_{m_{l_{M}}} d\sigma_2 \cdots d\sigma_N$$
  
where again $l_p >1$.   Note that $\int \prod_{i=1}^n[(\sigma_{l_p})_i^{m_i}] d\sigma_{l_p} =0$ unless all the integers $m_i$ are even and thus $I=0$ unless the integrand is a product of squares.  The fact that $I\ge 0$ shows the very simple Griffiths first inequality, $E f\ge 0$, even with ferromagnetic interaction.  

It remains to show that for $f \in \mathcal{E}$, $e^{t\Delta}f$ is a limit of functions in $\mathcal{E}$.
To show this we use a theorem of Paul Chernoff \cite {PC} to approximate $e^{t\Delta_i}$.  The operator $e^{t\Delta}$ in $\mathbb{R}^n$ is an integral operator with kernel $K_t(x,y) = (4\pi t)^{-n/2} e^{-|x-y|^2/4t}$.  Intuition tells us that a good small time approximation to $e^{t\Delta_0}$ where $\Delta_0$ is the spherical Laplacian on $\mathbb{S}^{n-1}$ should be the operator $U(t)$ given by 
$$U(t)f(\sigma) = c(t) \int e^{-|\sigma - \sigma'|^2/4t} f(\sigma') d\sigma'$$
where $d\sigma'$ is normalized Lebesgue measure on $\mathbb{S}^{n-1}$ and $c(t)$ is chosen so that $U(t) 1 = 1$.  
  
We can calculate $c(t)$ for $t > 0$ and small.  We find 
$$c(t) = A_{n-1} (4 \pi t)^{-(n-1)/2} (1 + O(t))$$  where $A_{n-1}$ is the area of $\mathbb{S}^{n-1}$ ($A_{n-1} = 2\pi^{n/2} / \Gamma(n/2)$). 
Clearly $ ||U(t)||_p = 1$ for $p = 1,\infty$ which gives $||U(t)||_p = 1$ for all $p \in [1,\infty]$ by the Riesz convexity theorem.  Here we also use $U(t) 1 = 1$ to get equality.

We use  Chernoff's theorem:

\bt \cite {PC} \label {Chernoff}
Suppose $U(t)$ is a strongly continuous family of contractions for $t\ge 0$ on a Banach space $B$ with $U(0) = I$.  Suppose $A$ is the generator of a strongly continuous contraction semigroup on $B$ such
that $(U(t) - I)t^{-1}f$ converges strongly to $Af$ on a core for $A$.  Then $U(t/m)^m$ converges strongly to $e^{tA}$ as $m \to \infty$.
\et

We use this for $A= \Delta_0$ , $B = L^2(\mathbb{S}^{n-1})$.  Note that the set of all $f = e^{s\Delta_0}\psi$ for some $s>0$ and some $\psi \in L^2$ is a core for $\Delta_0$ and thus we need only prove convergence on smooth functions.   Note also that one need not show $\nabla U(t/n)^n f \rightarrow \nabla e^{t\Delta_0} f$ since we have $d/dt (g,e^{t\Delta} f) = (\Delta g,e^{t\Delta} f)$. Thus consider 
$$((U(t) - I)t^{-1}f)(\sigma) = c(t) \int_{\mathbb{S}^{n-1}} e^{-|\sigma - \sigma'|^2/4t} (f(\sigma') - f(\sigma))t^{-1} d\sigma'$$ for smooth $f$.  Fix $\sigma$ and note that integration over $|\sigma' - \sigma| > \delta $ for any positive $\delta$ gives a negligible contribution so we can assume that $f$ has small support around $\sigma$ and that we integrate only over a small region of the sphere around $\sigma$.  We use coordinates on the sphere around $\sigma$ so that for $\sigma'\cdot \sigma > 0  , \sigma' = \sqrt{1-|x|^2} \sigma + x$ with $x$ in the $(n-1)$ - plane perpendicular to $\sigma$.  We see that 
$$|\sigma' - \sigma|^2 = |x|^2 ( 1 + |x|^4( \sqrt {1-|x|^2} +1)^{-2})$$
Without loss of generality (using the fact that $U(t)$ commutes with rotations) we take the $n$th component of $\sigma$ equal to $1$ so that we can write $x = (x_1,\cdots,x_{n-1})$, dropping $x_n = 0.$
In this coordinate system, at the point $\sigma$ we have $d\sigma'_1 \cdot d\sigma'_2 = \sum_{i,j} g_{i,j}(dx_i)_1(dx_j)_2$ with $g_{i,j} = \delta_{i,j} + x_ix_j/(1-|x|^2)$.  Thus $\det(g_{i,j}) = (1-|x|^2)^{-1} $ and 

$$I = ((U(t) - I)t^{-1}f)(\sigma) = $$ $$c(t) \int_{|x| <1/2}  e^{-|x|^2( 1 + |x|^4( \sqrt {1-|x|^2} +1)^{-2})/4t}) (\tilde f(x) - \tilde f(0))t^{-1}\frac{dx}{\sqrt{1-|x|^2}} + O(e^{-1/20t})$$
where $dx$ is Lebesgue measure on the plane perpendicular to $\sigma$ and $\tilde f(x) = f(\sigma')$.  It is easy to see that $c(t) \int_ {|x| < 1/2} |x|^3 e^{-|x|^2/4t} t^{-1} dx  \le C\sqrt{t}$ so that since odd functions of $x$ give zero integrals

$$I = c(t) \sum_ {i,j = 1}^{n-1} \frac{1}{2} \partial_i \partial_j \tilde f(0) \int _{|x| < 1/2} e^{-|x|^2( 1 + |x|^4( \sqrt {1-|x|^2} +1)^{-2})/4t}) x_ix_j t^{-1} \frac{dx}{\sqrt{1-|x|^2}} + O(\sqrt t) = $$
$$  c(t) \sum_ {i,j = 1}^{n-1}\frac{1}{2} \partial_i \partial_j \tilde f(0) \int _{|x| < 1/2} e^{-|x|^2/4t} x_ix_j t^{-1} dx+ O(\sqrt t) =$$ 
$$c(t) \frac{1}{2}\partial \cdot\partial \tilde f(0) \int_{|x|< 1/2}  e^{-|x|^2/4t} ( |x|^2/t(n-1)) dx + O(\sqrt t) .$$
We can integrate over all of $\mathbb{R}^{n-1}$ with a negligible error.  Substituting  $y = x/\sqrt {2t}$ we obtain 
$$ I = \frac{1}{2}\partial \cdot \partial \tilde f(0) c(t)(\sqrt {2t})^{n-1} \int e^{-|y|^2/2} (2|y|^2/(n-1)) dy  + O(\sqrt t) = $$
$$ \partial \cdot \partial \tilde f(0) c(t) (\sqrt{4\pi t})^{n-1} + O(\sqrt t).$$
Since $c(t) (\sqrt {4\pi t})^{n-1} =  1 + O(t)$
$$I = \partial \cdot \partial \tilde f(0) + O(\sqrt t).$$

It is easy to calculate the inverse of $g_{i,j}$.  We obtain $g^{i,j} = \delta_{i,j}  - x_ix_j$, so that in this coordinate system $\Delta_0 = \sum _{i,j} (\sqrt g)^{-1}  \partial_i \sqrt g g^{i,j} \partial_j$ ($g = \det (g_{i,j}) = (1-|x|^2)^{-1}$) or
$$\Delta_0 = -(n-1)x \cdot \partial + \sum_{i,j = 1}^{n-1} (\delta_{i,j} - x_i x_j)\partial_i \partial_j.$$

Thus according to Theorem \ref{Chernoff} we have $U(t/n)^n \rightarrow e^{t\Delta_0}$.  It remains to show that for $f \in \mathcal{E}, U(t)f$ is a limit of functions in $\mathcal{E}$
or more explicitly we must show that $$\int e^{-\sum_{i=1}^N |\sigma_i - \sigma_i '|^2/4t} f(\sigma'_1,\cdots,\sigma'_N) d\sigma'_1,\cdots,d\sigma'_N$$ is a limit of functions in $\mathcal{E}$ .  Since $e^{-|\sigma_i - \sigma_i'|^2/4t} = e^{-1/2t} e^{\sigma_i\cdot\sigma_i'/2t}$, expanding the second exponential in a power series we see that if suffices to show that $$\int (\sigma_1\cdot \sigma_1')^{n_1}\cdots(\sigma_N\cdot \sigma_N')^{n_N}  f(\sigma'_1,\cdots,\sigma'_N) d\sigma'_1,\cdots,d\sigma'_N \in \mathcal{E}.$$  If $f$ is a sum of terms of the form $c \prod _{i < j} (\sigma'_i \cdot \sigma'_j)^{n_{i,j}}$ with $c>0$ it thus suffices to show that the more general integral
$$J = \int \prod_{j_1} (\sigma_1\cdot \sigma_{j_1}')^{m_{j_1}}\cdots\prod_{j_N}(\sigma_N\cdot \sigma_{j_N}')^{m_{j_N}}  \prod_{i < j} (\sigma'_i \cdot \sigma'_j)^{n_{i,j}} d\sigma'_1,\cdots,d\sigma'_N \in \mathcal{E}.$$ 
Let us do the integral over $\sigma'_1$, namely an integral of the form
$$I = \int \prod_{1\le k}(\sigma'_1 \cdot \sigma_k)^{n_k} \prod _{1 < j} (\sigma'_1 \cdot \sigma'_j)^{n_{1,j}} d\sigma'_1.$$
We convert this to a Gaussian integral
$$ I = C\int_{\mathbb{R}^n}e^{-|x|^2/2}\prod_{1\le k}(\sigma_k \cdot x)^{n_k} \prod _{1 < j} (x \cdot \sigma'_j)^{n_{1,j}} dx/(\sqrt{2\pi})^n$$
where $C^{-1} =A_{n-1} \int_0^\infty s^{(\sum n_k + n-1 + \sum_j n_{1,j})} e^{-s^2/2} ds/(\sqrt{2\pi})^{-n}$ and $A_{n-1}$  is the area of $\mathbb{S}^{n-1}$. We now use Leon Isserlis' theorem \cite {I} (better known as Wick's theorem \cite {W} in the physics literature) to compute this integral.  The theorem states in particular that 
$$\int_{\mathbb{R}^n}e^{-|x|^2/2} \prod_{i=1}^L (a_i \cdot x) dx/(\sqrt{2\pi})^n = \sum_{p \in P_L} \prod_{(i,j) \in p}(a_i \cdot a_j)$$
where $P_L$ is the set of pairings of $\{1,\cdots,L\}$, i.e. the distinct ways of partitioning $\{1,\cdots,L\}$ into pairs $(i,j)$.  The product is over the pairs contained in the partition $p \in P_L$.  If $L$ is odd, the integral is $0$. Thus either $I=0$ or $I$ is a sum of terms, each a product of a function in $\mathcal{E}$ and a product $\prod_ {1 < i <j} (\sigma'_i \cdot \sigma'_j)^{l_{i,j}}$.  Thus in the integral $J$, after doing the $\sigma'_1$ integral we have a sum of terms involving a factor from $\mathcal{E}$ times an integral of the same form as $J$ except $\sigma'_1$ is omitted.  Iteration shows that $J\in \mathcal{E}$ and completes the demonstration.
\epf
 $\blacksquare$

\section{Gaussian ferromagnetic spins}

In this section, using semigroup techniques, we prove the Griffiths inequalities for $n$ dimensional spins in a ferromagnetic Gaussian measure $Z^{-1} e^{- \sum _{i,j =1}^N f_{i j} x_i\cdot x_j/2}$ where the matrix $F= (f_{i j})$ is positive definite with $f_{ij} \le 0$ for $i\ne j$.  There is a much simpler proof (see \cite{GS} for example) which proves the stronger Ginibre inequalities \cite{G} by a simple change of variable. But again we give a brief outline of a (longer) proof which we hope might spur additional research.  

\begin{proposition}
Define the expectation 
$$E f = \int f(x_1,x_2,\cdots.x_N) d\mu(x_1,\cdots,x_N),$$
$$d\mu(x_1,\cdots,x_N ) =  Z^{-1} e^{- \sum _{i,j =1}^N f_{i j} x_i\cdot x_j/2}dx_1\cdots dx_N$$
where the matrix $F= (f_{i j})$ is symmetric and positive definite with $f_{ij} \le 0$ for $i\ne j$.  $Z$ is defined so that $\mu$ is a probability measure.  The variables $x_i \in \mathbb{R}^n$. Let $\mathcal{E}$ be the cone of non-negative linear combinations of multinomials $\prod_{i \le j}  (x_i\cdot x_j)^{n_{i,j}}$ with integers $n_{i,j} \ge 0$.  Then if $f,g \in \mathcal{E}$
$$ Ef \ge 0, Efg \ge Ef Eg.$$
\end{proposition}

We give a brief outline of the semigroup proof:  We define the self adjoint operator $A$ in $L^2(e^{-Q} dx)$ with $Q = \sum f_{i j} x_i\cdot x_j/2$ by the formula 
   $$\int f A g e^{-Q} dx = -\int \nabla f \cdot \nabla g e^{-Q} dx.$$
 We calculate $A = \Delta - \nabla Q \cdot \nabla$.  ($A$ is related to the Ornstein-Uhlenbeck process. The unitary transform of $A$, $e^{-Q/2} A e^{Q/2} $, is the negative of a harmonic oscillator Hamiltonian, $ H = -\Delta + |Fx|^2/4 - n \text{tr} F$. ) We use the Trotter formula \cite {T} which gives $\lim_{m \to \infty} (e^{t\Delta/m} e^{-t\nabla Q \cdot \nabla/m})^m = e^{tA}$.  $e^{t\Delta}$ preserves $\mathcal{E}$ as does $e^{-t\nabla Q \cdot \nabla}$ which just takes $f(x)$ to $f(e^{-tF}x)$.  The ferromagnetic nature of $F$ guarantees that the matrix elements of $e^{-tF}$ are non-negative (proof: $e^{-tF} = \lim_{m \to \infty} (I-tF/m)^m$) and thus $e^{-tF}$  preserves $\mathcal{E}$.  Finally Isserlis' theorem shows that the resulting integrals are non-negative (here we use that $F^{-1} = \int_0^\infty e^{-tF} dt$ has non-negative entries).  
   

DEPARTMENT OF MATHEMATICS, UNIVERSITY OF VIRGINIA, CHARLOTTESVILLE, VA 22904.\\
\emph{E-mail address: iwh@virginia.edu}


\begin{thebibliography}{99}
\bibitem{Gr1} Griffiths, R., \emph{Correlations in Ising ferromagnets, I}, J. Math. Phys., $\mathbf{8}$, (1967), 478--483.
\bibitem{Gr2} Griffiths, R., \emph{Correlations in Ising ferromagnets, II}, J. Math. Phys., $\mathbf{8}$, (1967), 484--489.
\bibitem{KS} D. Kelly, S. Sherman, \emph{General Griffiths' inequalities on correlations in Ising ferromagnets}, J. Math. Phys., $\mathbf{9}$, (1968), 466--484.
\bibitem{G} Ginibre, J., \emph{General formulation of Griffith's inequalities}, Comm. Math. Phys., $\mathbf{16}$, (1970), 310--328.
\bibitem{MP} Monroe, J., Pearce, P., \emph{Correlation inequalities for vector spin models}, J. Stat. Phys., $\mathbf{21}$, 6, (1979), 615--633.
\bibitem{P} Pitt, L., \emph{A Gaussian correlation inequality for symmetric convex sets}, Ann. Prob., $\mathbf{5}$, (3), (1977), 470--474.
\bibitem{HP} Herbst, I. and Pitt, L., \emph{Diffusion equation techniques in stochastic monotonicity and positive correlations}, Prob, Theor. Rel. Fields, $\mathbf{87}$, (3), (1991), 275--312.
\bibitem{PC} Chernoff, P. \emph{Note on product formulas for operator semigroups}, J. Funct. Anal., $\mathbf{2}$, (1968), 238--242.
\bibitem{I} Isserlis, L., \emph{On a formula for the product-moment coefficient of any order of a normal frequency distribution in any number of variables} Biometrika, $\mathbf {12}$, (1-2), (1918), 134--139. 
\bibitem{W} Wick, G. C., \emph{The evaluation of the collision matrix}, Phys. Rev., $\mathbf{80}$, (2), (1950), 268--272. 
\bibitem{B} Baumgartner, B. \emph{Griffiths inequalities for noninteracting N-vector (classical Heisenberg) models and applications to interacting systems}, J. Stat. Phys. $\mathbf{32}$, (3),(1983), 615--625.
\bibitem{GS} Sylvester, G., \emph{The Ginibre inequality}, Comm. Math. Phys., $\mathbf{73}$, (1980), 105--114.
\bibitem{T} Trotter, H. F. \emph{On the product of semigroups of operators}, Proc. AMS, $\mathbf{10}$ (1959), 545--551.
\end{thebibliography}
\end{document}